\documentclass[iop]{emulateapj}

\usepackage{natbib,graphicx,amsmath}

\shorttitle{SPECIAL RELATIVISTIC HYDRODYNAMICS}
\shortauthors{HWANG \& NOH}

\newcommand{\bea}{\begin{eqnarray}}
\newcommand{\eea}{\end{eqnarray}}

\begin{document}

\title{Special relativistic hydrodynamics with gravitation}
\author{Jai-chan Hwang${}^{1}$, Hyerim Noh${}^{2}$}
\address{${}^{1}$Department of Astronomy and Atmospheric Sciences,
                 Kyungpook National University, Daegu, Korea \\
         ${}^{2}$Korea Astronomy and Space Science Institute,
                 Daejon, Korea
                 }

\begin{abstract}

The special relativistic hydrodynamics with weak gravity is hitherto unknown in the literature. Whether such an asymmetric combination is possible was unclear. Here, the hydrodynamic equations with Poisson-type gravity considering fully relativistic velocity and pressure under the weak gravity and the action-at-a-distance limit are consistently derived from Einstein's general relativity. Analysis is made in the maximal slicing where the Poisson's equation becomes much simpler than our previous study in the zero-shear gauge. Also presented is the hydrodynamic equations in the first post-Newtonian approximation, now under the {\it general} hypersurface condition. Our formulation includes the anisotropic stress.

\end{abstract}

\keywords{hydrodynamics, gravitation, relativity}

\section{Introduction}

Dealing with the full machinery of Einstein's gravity in astrophysical situations is often a formidable task both conceptually and practically, especially in the numerical study. It is desirable to have approximation methods if the situation allows. One such a well known approximation is the post-Newtonian (PN) method (Chandrasekhar 1965; Chandrasekhar \& Nutku 1969; Chandrasekhar \& Esposito 1970; Weinberg 1972; Blanchet, Damour \& Sch\"afer 1990; Shibata \& Asada 1995; Asada, Shibata \& Futamase 1996; Asada \& Futamase 1997; Takada \& Futamase 1999; Hwang, Noh \& Puetzfeld 2008, HNP hereafter). The PN approximation takes into account of the effect of Einstein's gravity as correction terms in Newtonian equations, thus weakly relativistic but fully nonlinear. In this work, we will present the first PN (1PN) hydrodynamic equations in the {\it general} temporal gauge (hypersurface or slicing) condition, Section \ref{sec:PN}.

Besides the 1PN approximation, our main objective in this work is to present another approximation which is the special relativistic hydrodynamics in the presence of weak self-gravity with the action-at-a-distance limit. As we take into account the fully relativistic pressure, anisotropic stress and velocity, this approximation is complementary to the 1PN method and full nonlinear as well; compared with the 1PN approximation our gravity and the action-at-a-distance conditions are weaker. Our equations will reveal how the relativistic pressure, anisotropic stress and velocity gravitate, Section \ref{sec:Summary}. The equations will be derived consistently from Einstein's equation, Section \ref{sec:Derivation}.

Einstein's gravity causes the result to depend on the temporal gauge choice. Previously we studied it in the zero-shear gauge (ZSG) (Hwang \& Noh 2013c, HNc hereafter; Hwang et al.\ 2016). Here, we take the maximal slicing (MS) which sets the trace of extrinsic curvature equal to be zero (Smarr \& York 1978). The MS causes a difference {\it only} in the Poisson's equation which becomes much simpler than the previous one in the ZSG and suitable for numerical study. Here we include the anisotropic stress.

We use the terms temporal gauge, hypersurface and slicing condition interchangeably. In cosmology the MS corresponds to the uniform-expansion gauge taking the perturbed part of the expansion scalar in the normal frame (which is the same as the perturbed part of trace of extrinsic curvature with a minus sign) to be zero. The gauge degrees of freedom are completely fixed in this slicing (as well as the ZSG) together with our (spatial) gauge conditions imposed in the spatial part of the metric, and each variable in this gauge has a unique corresponding gauge-invariant combination to the fully nonlinear order (Bardeen 1988; Noh \& Hwang 2004; Hwang \& Noh 2013a, HNa hereafter).

Previously we presented a fully nonlinear and exact cosmological perturbation theory in the absence of spatially transverse-tracefree perturbation (HNa). Although the formulation was made in the cosmological context it is applicable to the non-expanding background by setting the scale factor equal to one, and letting the background Robertson-Walker metric to be the Minkowsky one. We have shown that in the $c$-goes-to-infinity limit our perturbation formulation of Einstein's gravity properly reproduce the Newtonian hydrodynamics in two gauge conditions (Hwang \& Noh 2013b, HNb hereafter): these are the ZSG and the uniform-expansion gauge (the MS in Minkowsky background). The 1PN equations are also properly recovered without fixing the temporal gauge condition (Noh \& Hwang 2013).

As we have a curious difference of the presence/absence of the $3p/c^2$ term in the Poisson's equation depending on the gauge [compare Equations (\ref{Poisson-MS}) and (\ref{Poisson-ZSG})], we pay attention to it in the corresponding PN formulation (Section \ref{sec:PN}) as well as in the gauge transformation to the linear order (Section \ref{sec:GT}).

In the Appendix A we present the special relativistic hydrodynamics in the Minkowsky spacetime now including the anisotropic stress; as far as we are aware the case with the stress is not known in the literature. In the Appendix B we present the fully nonlinear and exact perturbation equations in a gauge ready (i.e., without taking the temporal gauge) form ignoring the transverse-tracefree type spatial metric in nonexpanding background; this is also new in the literature.

Section \ref{sec:Summary} is a summary of the newly proposed special relativistic hydrodynamic formulation with gravity. Section \ref{sec:Derivation} derive the formulation from full Einstein's gravity in the two gauge conditions. Section\ \ref{sec:PN} presents the complementary 1PN equations without taking the slicing condition.

For readers interested in numerical implementation, here we summarize the fundamental sets of equations derived in this work. Equations (\ref{continuity-practice})-(\ref{Poisson-MS}) are the complete set of equations for special relativistic hydrodynamic with gravity in the MS; for more generally valid form see Equations (\ref{continuity})-(\ref{M-conservation}).  An alternative expression is Equations (\ref{continuity-SR-practice})-(\ref{M-conservation-SR-practice}) and (\ref{Poisson-MS-E}). In the post-Newtonian approximation Equations (\ref{1PN-0})-(\ref{1PN-5}) are the complete set with the general gauge condition in Equation (\ref{1PN-gauge}).

\begin{widetext}
%
%
\section{Special relativistic hydrodynamics with gravitation: summary}
                                             \label{sec:Summary}

The special relativistic extension of the mass conservation, the energy conservation, the momentum conservation, and the Poisson's equations, respectively, in the MS is
\bea
   & &
       {d \overline \varrho \over dt}
       + \overline \varrho
       \left( \nabla \cdot {\bf v} + {d \over dt} \ln{\gamma} \right)
       = 0,
   \label{continuity-practice} \\
   & &
       {d \varrho \over dt}
       + \left( \varrho + {p \over c^2} \right)
       \left( \nabla \cdot {\bf v} + {d \over dt} \ln{\gamma} \right)
       = - {1 \over c^2} \Pi_i^j \nabla_j v^i
       - {1 \over c^4} \Pi_{ij} v^i \dot v^j,
   \label{E-conservation-practice} \\
   & &
       {d {\bf v} \over dt}
       =
       \nabla \Phi
       - {1 \over \varrho + p/c^2} {1 \over \gamma^2}
       \left( \nabla p
       + {1 \over c^2} {\bf v} \dot {p} \right)
   \nonumber \\
   & & \qquad
       + {1 \over \varrho + p/c^2} {1 \over \gamma^2} \left\{
       - \Pi^j_{i,j}
       + {1 \over c^2} \left[ {\bf v} \left( \Pi_j^k v^j \right)_{,k}
       - {1 \over \gamma^2} \left( \Pi_{ij} v^j \right)^{\displaystyle\cdot} \right]
       + {1 \over c^4} {\bf v}
       \left( \Pi_{jk} v^j v^k \right)^{\displaystyle\cdot} \right\},
   \label{M-conservation-practice} \\
   & &
       \Delta \Phi
       + 4 \pi G \left( \varrho + 3 {p \over c^2} \right)
       = - {8 \pi G \over c^2} \left[
       \left( \varrho + {p \over c^2} \right) \gamma^2 v^2
       + \Pi^i_i \right],
   \label{Poisson-MS}
\eea
with
\bea
   & &
       {d \over dt} \equiv {\partial \over \partial t}
       + {\bf v} \cdot \nabla, \quad
       \gamma = {1 \over \sqrt{ 1 - {v^2/ c^2}}},
   \label{gamma}
\eea
where $\varrho \equiv \overline \varrho (1 + \Pi/c^2)$, $p$ and ${\bf v}$ are the density, pressure and velocity, respectively, and $\gamma$ is the Lorentz factor with $v^2 \equiv v^k v_k$; $\overline \varrho$ is the rest mass density and $\overline \varrho \Pi$ is the internal energy density; $\Pi_{ij}$ is the anisotropic stress, see Equation (\ref{Pi_ij}).

These are complete (closed) system; the $p$ and $\Pi_{ij}$ are provided by the equations of state depending on the medium we are considering. Equation for $\ln{\gamma}$ follows from Equation (\ref{M-conservation-practice}), and is presented in Equation  (\ref{M-conservation-3}). Notice that the effect of gravity on the dynamics in Equation (\ref{M-conservation-practice}) is rather conventional. Except for the gravity Equations (\ref{continuity-practice})-(\ref{M-conservation-practice}) are the same as in the special relativistic hydrodynamics (see the Appendix A); For conventional forms often used in the numerical studies see Equations (\ref{continuity-SR})-(\ref{M-conservation-SR}). Equations (\ref{continuity-practice}) and (\ref{E-conservation-practice}) give
\bea
   & &
       \overline \varrho {d \Pi \over dt}
       + p \left( \nabla \cdot {\bf v} + {d \over dt} \ln{\gamma} \right)
       = - \Pi_i^j \nabla_j v^i
       - {1 \over c^2} \Pi_{ij} v^i \dot v^j.
   \label{Pi-equation-practice}
\eea

Our metric convention is
\bea
   & &
       ds^2 = - \left( 1 - {2 \Phi \over c^2} \right) c^2 d t^2
       - 2 \chi_i c dt d x^i
       + \left( 1 + {2 \Psi \over c^2} \right) \delta_{ij} d x^i d x^j.
   \label{metric}
\eea
The index of $\chi_i$ is raised and lowered by $\delta_{ij}$ as the metric. Keeping $\chi^i$ in the metric is essentially important in the MS analysis; see below Equation (\ref{Phi-Psi}). We {\it ignored} the transverse-tracefree perturbation in the spatial metric, and imposed spatial gauge conditions so that the spatial part of the metrics becomes simple as in Equation (\ref{metric}) to the fully nonlinear orders (Bardeen 1988; Noh \& Hwang 2004; HNa); in the spatial part of the metric tensor
\bea
   & &
       g_{ij} = \left( 1 + {2 \Psi \over c^2} \right) \delta_{ij}
       + 2 \overline \gamma_{,ij}
       + C^{(v)}_{i,j} + C^{(v)}_{j,i}
       + 2 C^{(t)}_{ij},
\eea
with $C^{(v)i}_{\;\;\;\;\;\;,i} \equiv 0$ and $C^{(t)j}_{\;\;\;\;\;i,j} \equiv 0 \equiv C^{(t)i}_{\;\;\;\;\;i}$, we imposed the spatial gauge conditions as
\bea
   & &
       \overline \gamma \equiv 0 \equiv C^{(v)}_i,
   \label{spatial-gauge}
\eea
and set $C^{(t)}_{ij} = 0$ as we ignore the transverse-tracefree part of the metric. We take the same spatial gauge in the 1PN approximation, see Equation (\ref{metric-1PN}) (HNP). Although we ignore the transverse-tracefree part of the metric we keep the corresponding part of the anisotropic stress. Thus we are considering situations where the tensor part (gravitational waves) of the metric is not significantly excited; for example, in the PN approximation the $C^{(t)}_{ij}$ part appears from the 2.5PN order (Chandrasekhar \& Esposito 1970).

The $\Psi$ does not affect the fluid motions and is determined by
\bea
   & &
       \Delta \Psi + 4 \pi G \varrho
       = - {4 \pi G \over c^2} \left[ \left(
       \varrho + {p \over c^2} \right) \gamma^2 v^2
       + \Pi^i_i \right].
   \label{Poisson-Psi}
\eea
Thus, with relativistic pressure, anisotropic stress or velocity, $\Psi$ in general differs from $\Phi$. Notice how the relativistic pressure, anisotropic stress and velocity gravitate. We may call Equations (\ref{Poisson-MS}) and (\ref{Poisson-Psi}) the extended Newtonian and the PN Poisson's equations, respectively, where $\Phi$ and $\Psi$ may be termed as the Newtonian and the PN gravitational potentials, respectively; in the Newtonian limit $\Psi$ does not have role in Newtonian dynamics, but we have $\Psi = \Phi$.

In the case of the ZSG which sets the longitudinal part of $\chi_i$ equal to zero, the only difference appears in the Newtonian Poisson's equation. Instead of Equation (\ref{Poisson-MS}) we have
\bea
   & &
       \Delta \Phi
       + 4 \pi G \varrho
       = - {8 \pi G \over c^2} \left[
       \left( \varrho + {p \over c^2} \right)
       \gamma^2 v^2 + \Pi^i_i \right]
       + {12 \pi G \over c^2}
       \Delta^{-1} \nabla_i \nabla_j
       \left[ \left( \varrho + {p \over c^2} \right)
       \gamma^2 {v^i v^j} + \Pi^{ij} \right],
   \label{Poisson-ZSG}
\eea
and Equations (\ref{continuity-practice})-(\ref{M-conservation-practice}) and (\ref{Poisson-Psi}) remain the same (Hwang et al.\ 2016). Notice the complication involving the inverse Laplacian operator $\Delta^{-1}$ and the absence of $3 p/c^2$ term in Equation (\ref{Poisson-ZSG}) compared with Equation (\ref{Poisson-MS}) in the MS.

In order to derive the above equations we {\it assume} the weak gravity and the action-at-a-distance conditions
\bea
   & &
       {\Phi \over c^2} \ll 1, \quad
       {\Psi \over c^2} \ll 1, \quad
       \gamma^2 {t_\ell^2 \over t_g^2} \ll 1,
   \label{assumptions}
\eea
where $t_g \sim 1/\sqrt{G \varrho}$ and $t_\ell \sim \ell/c \sim 2 \pi /(kc)$ are gravitational time scale and the light propagating time scale of a characteristic length scale $\ell$, respectively; $k$ is the wave number with $\Delta = - k^2$. The action-at-a-distance condition implies that we keep the action-at-a-distance nature of Newtonian theory. In our approximation presented above we {\it ignore} these dimensionless quantities compared with order unity.

We have
\bea
   & &
       \gamma^2 {t_\ell^2 \over t_g^2}
       \sim \gamma^2 {G \varrho \ell^2 \over c^2}
       \sim \gamma^2 {G M \over \ell c^2}
       \sim {\Phi \over c^2}
       \sim {\Psi \over c^2},
\eea
where in the last two steps we used our special relativistic Poisson's equations in Equations (\ref{Poisson-MS}), (\ref{Poisson-Psi}) and (\ref{Poisson-ZSG}); $M$ is the characteristic mass scale. This estimate shows that all the conditions in Equation (\ref{assumptions}) are of the same order. Thus, even for ultra-relativistic velocity ($\gamma \gg 1$) the action-at-a-distance condition is comparable to the weak gravity conditions due to the relativistic boost corrections in the Poisson's equations. Still, in such an ultra-relativistic situation we can make an argument that unless we have steep gradient in the gravitational potential the gravity term is negligible in the hydrodynamic equations, see below Equation (\ref{M-conservation-4}); without the steep gradient in $\Phi$, $\nabla \Phi$ term in Equation (\ref{M-conservation-practice}) is negligible compared with the convective term in the left-hand-side due to the weak gravity condition. This is true despite the fact that in Equations (\ref{Poisson-MS}), (\ref{Poisson-Psi}) and (\ref{Poisson-ZSG}) the Lorentz boosted density and pressure in the right-side-sides of these equations dominate the gravity. In the presence of steep gradients of $\Phi$ and $\Psi$ we have a more general extended hydrodynamic conservation equations derived in Equations (\ref{continuity})-(\ref{M-conservation}), while the Poisson's equations in Equations (\ref{Poisson-MS}), (\ref{Poisson-Psi}) and (\ref{Poisson-ZSG}) remain the same.
The gravity might have non-negligible roles in situations with mildly relativistic velocity (caused by non-gravitational means) with $\gamma \sim 1$.

\subsection{Newtonian limit}

By taking $c$-goes-to-infinity limit we recover the Newtonian hydrodynamic equations in both gauges (HNb). We have
\bea
   & &
       {\partial \overline \varrho \over \partial t}
       + \nabla \cdot \left( \overline \varrho {\bf v} \right)
       = 0,
   \\
   & &
       {\partial {\bf v} \over \partial t}
       + {\bf v} \cdot \nabla {\bf v}
       = \nabla \Phi
       - {1 \over \varrho} \left( \nabla p
       + \nabla_j \Pi^j_i \right),
   \\
   & &
       \Delta \Phi + 4 \pi G \varrho = 0.
\eea
Even in the Newtonian context, the internal energy equation  (\ref{Pi-equation-practice}) in the c-infinity limit becomes
\bea
   & &
       \overline \varrho {d \Pi \over dt}
       + p \nabla \cdot {\bf v}
       = - \Pi_i^j \nabla_j v^i.
\eea
Including the internal energy the energy conservation equation (\ref{E-conservation-practice}) becomes
\bea
   & &
       {d \varrho \over dt}
       + \left( \varrho + {p \over c^2} \right)
       \nabla \cdot {\bf v}
       = - {1 \over c^2} \Pi_i^j \nabla_j v^i.
\eea

\subsection{Slow-motion and Newtonian stress limit}

Now, for non-relativistic velocity ($v^2/c^2 \rightarrow 0$) and Newtonian anisotropic stress, but still with fully relativistic pressure, Equations (\ref{continuity-practice})-(\ref{Poisson-MS}) become
\bea
   & &
       {d \overline \varrho \over dt}
       + \overline \varrho \nabla \cdot {\bf v}
       = {\overline \varrho \over \varrho + p/c^2}
       {1 \over c^2} {\bf v} \cdot \nabla p,
   \label{continuity-NR} \\
   & &
       {d \varrho \over dt}
       + \left( \varrho + {p \over c^2} \right)
       \nabla \cdot {\bf v}
       = {1 \over c^2} {\bf v} \cdot \nabla p
       - {1 \over c^2} \Pi_i^j \nabla_j v^i,
   \label{E-conservation-NR} \\
   & &
       {d {\bf v} \over dt}
       = \nabla \Phi
       - {1 \over \varrho + p/c^2}
       \left( \nabla p
       + {1 \over c^2} {\bf v} \dot p
       + \nabla_j \Pi_i^j \right),
   \label{M-conservation-NR} \\
   & &
       \Delta \Phi
       + 4 \pi G \left( \varrho + 3 {p \over c^2} \right)
       = 0,
   \label{Poisson-MS-NR}
\eea
and Equation (\ref{Poisson-Psi}) becomes
\bea
   & &
       \Delta \Psi + 4 \pi G \varrho = 0.
   \label{Poisson-Psi-NR}
\eea
In the ZSG, instead of Equation (\ref{Poisson-MS-NR}) we have (HNc)
\bea
   & &
       \Delta \Phi + 4 \pi G \varrho
       = 0.
   \label{Poisson-ZSG-NR}
\eea
The only difference between the two gauges is the presence or absence of $3p/c^2$ term in the Poisson's equation. Equations (\ref{continuity-NR}) and (\ref{E-conservation-NR}) give
\bea
   & &
       \overline \varrho {d \Pi \over dt}
       + p \nabla \cdot {\bf v}
       = {p/c^2 \over \varrho + p/c^2} {\bf v} \cdot \nabla p
       - \Pi_i^j \nabla_j v^i.
\eea

\subsection{Static limit}

In a static equilibrium situation with all time derivatives vanishing, Equation (\ref{M-conservation-practice}) leads to
\bea
   & &
       \nabla_i \Phi = {1 \over \varrho + p/c^2}
       \left( \nabla_i p + \nabla_j \Pi_i^j \right).
\eea
Equations (\ref{Poisson-MS}) and (\ref{Poisson-Psi}) become
\bea
   & &
       \Delta \Phi = - 4 \pi G \left( \varrho + 3 {p \over c^2} \right),
   \label{Poisson-MS-static} \\
   & &
       \Delta \Psi = - 4 \pi G \varrho,
\eea
where we used Equation (\ref{Pi_ii}).

Further assuming a spherical symmetry, we have
\bea
   & &
       {dp \over dr}
       = - {4 \pi G \over r^2} \left( \varrho + {p \over c^2} \right)
       \int_0^r \left( \varrho + {3 p \over c^2} \right) r^2 dr,
   \label{static-case}
\eea
which is the well known Tolman-Oppenheimer-Volkoff relation in the weak gravity limit (Tolman 1939, Oppenheimer \& Volkoff 1939). The presence of $3p/c^2$-terms in the slow-motion limit of Equations (\ref{Poisson-MS}) and (\ref{Poisson-MS-NR}) are consistent with the historically known results (Whittaker 1935, McCrea 1951, Harrison 1965).

It is curious to notice, however, that this $3 p/ c^2$-term is absent in the ZSG: see Equations (\ref{Poisson-ZSG}) and (\ref{Poisson-ZSG-NR}).  In the ZSG, instead of Equation (\ref{Poisson-MS-static}), we have
\bea
   & &
       \Delta \Phi = - 4 \pi G \left( \varrho
       + {3 \over c^2} \Delta^{-1} \nabla_i \nabla_j \Pi^{ij} \right).
\eea
This makes the ZSG result inconsistent with the Tolman-Oppenheimer-Volkoff relation by missing $3p/c^2$ term in Equation (\ref{static-case}) which can be regarded as a serious disqualification in the spherically symmetric situation. At the moment the meaning of why we have such an inconsistent result in the ZSG is unclear to us; derivations in the ZSG in HNc and Hwang et al.\ (2016) are consistent with full Einstein's equations just as in the case for the MS performed in this work; the ZSG case is also derived in parallel below. This has motivated us to pay special attention to this difference (presence or absence of $3p/c^2$-term in the Poisson's equation depending on hypersurface) by analyzing the 1PN approximation and gauge transformation in perturbation theory: the difference is real, see paragraphs below Equation (\ref{1PN-gauge}) and Section \ref{sec:GT}.

%
%
\section{Derivation}
                                             \label{sec:Derivation}

Here, we derive Equations (\ref{continuity-practice})-(\ref{Poisson-MS}), (\ref{Poisson-Psi}) and (\ref{Poisson-ZSG}) from full Einstein's equations assuming Equation (\ref{assumptions}). There is no {\it a priori} reason for a formulation based on such assumptions should be possible. The proof based on full Einstein's equation will guarantee the consistency as a result. We will use the fully nonlinear and exact perturbation formulation of the perturbation equations summarized in the Appendix B.

In our derivation of the conservation equations, we apply the weak gravity condition, but {\it kept} ${\bf v} \cdot \nabla \Phi/c^2$ term compared with $\nabla \cdot {\bf v}$, and ${\bf v} {\bf v} \cdot \nabla \Phi/c^2$ term compared with ${\bf v} \cdot \nabla {\bf v}$, etc. (not in the combination with the anisotropic stress though), although these terms are practically negligible in our weak gravity condition, see below Equation (\ref{M-conservation-4}). We ignore $\dot \Psi /c^2$ term compared with unity. Thus, instead of Equations (\ref{continuity-practice})-(\ref{M-conservation-practice}), we will derive the following more general forms
\bea
   & &
       {d \overline \varrho \over dt}
       + \overline \varrho
       \left[ \nabla \cdot {\bf v} + {d \over dt} \ln{\gamma} + {1 \over c^2} {\bf v} \cdot \nabla \left( \Psi - \Phi \right) \right]
       = 0,
   \label{continuity} \\
   & &
       {d \varrho \over dt}
       + \left( \varrho + {p \over c^2} \right)
       \left[ \nabla \cdot {\bf v} + {d \over dt} \ln{\gamma} + {1 \over c^2} {\bf v} \cdot \nabla \left( \Psi - \Phi \right) \right]
       = - {1 \over c^2} \Pi^i_j v^j_{\;\;,i}
       - {1 \over c^4} \Pi_{ij} v^i \dot v^j,
   \label{E-conservation} \\
   & &
       {d {\bf v} \over dt}
       =
       \nabla \Phi
       - {1 \over c^2} {\bf v} {\bf v} \cdot \nabla \Phi
       + {v^2 \over c^2} \nabla \Psi
       - {1 \over \varrho + p/c^2} {1 \over \gamma^2}
       \left( \nabla p
       + {1 \over c^2} {\bf v} \dot {p} \right)
   \nonumber \\
   & & \qquad
       + {1 \over \varrho + p/c^2} {1 \over \gamma^2} \left\{
       - \Pi^j_{i,j}
       + {1 \over c^2} \left[ {\bf v} \left( \Pi_j^k v^j \right)_{,k}
       - {1 \over \gamma^2} \left( \Pi_{ij} v^j \right)^{\displaystyle\cdot} \right]
       + {1 \over c^4} {\bf v}
       \left( \Pi_{jk} v^j v^k \right)^{\displaystyle\cdot} \right\}.
   \label{M-conservation}
\eea
Equations (\ref{continuity}) and (\ref{E-conservation}) give
\bea
   & &
       \overline \varrho {d \Pi \over dt}
       + p \left[ \nabla \cdot {\bf v} + {d \over dt} \ln{\gamma}
       + {1 \over c^2} {\bf v} \cdot \nabla \left( \Psi - \Phi \right) \right]
       = - \Pi^i_j v^j_{\;\;,i}
       - {1 \over c^2} \Pi_{ij} v^i \dot v^j.
   \label{Pi-equation}
\eea
In this more general forms, the relativistic pressure, anisotropic stress and velocity in general demand the presence of two gravitational potentials, $\Phi$ and $\Psi$. Although these will be derived consistently from Einstein's gravity, we will argue that these additional terms are, unless we have steep gradients in $\Psi$ and $\Psi$, practically negligible, see below Equation (\ref{M-conservation-4}). By ignoring these terms, thus strictly applying the weak gravity conditions in Equation (\ref{assumptions}), we have Equations (\ref{continuity-practice})-(\ref{M-conservation-practice}).

We consider a general fluid energy-momentum tensor
\bea
   & &
       T_{ab} = \left( \varrho c^2 + p \right) u_a u_b + p g_{ab}
       + \pi_{ab},
   \label{Tab}
\eea
with $\mu \equiv \varrho c^2$. The anisotropic stress is introduced as
\bea
   & &
       \pi_{ij} \equiv \Pi_{ij},
   \label{Pi_ij}
\eea
where the indices of $\Pi_{ij}$ is raised and lowered by $\delta_{ij}$ as the metric. From $\pi^c_c \equiv 0$ we have
\bea
      \Pi^i_i = \Pi_{ij} {v^i v^j \over c^2}.
   \label{Pi_ii}
\eea
The fluid four-vector is introduced as
\bea
   & &
       u_i \equiv \gamma {v_i \over c}, \quad
       \gamma \equiv - n_c u^c
       = {1 \over \sqrt{ 1 - {v^k v_k
       \over c^2 (1 + 2 \Psi/c^2) }}},
   \label{v_i}
\eea
where $v_i$ is the fluid three-velocity measured by the Eulerian observer (HNa); the index of $v_i$ is raised and lowered by $\delta_{ij}$ as the metric; $n_a$ is the normal-frame four-vector with $n_i \equiv 0$. In the weak gravity approximation the Lorentz factor $\gamma$ becomes the one in Equation (\ref{gamma}).

The ADM (Arnowitt-Deser-Misner) momentum constraint in Equation (\ref{eq3}) gives
\bea
   & &
       {2 \over 3} \kappa_{,i}
       + c \left( {1 \over 2} \Delta \chi_i
       + {1 \over 6} \chi^j_{\;\;,ij} \right)
       = - {8 \pi G \over c^2}
       \left[ \left( \varrho + {p \over c^2} \right)
       \gamma^2 v_i + {1 \over c^2} \Pi_{ij} v^j \right].
   \label{mom-constraint}
\eea
We decompose $\chi_i$ into the longitudinal and transverse parts as $\chi_i \equiv \chi_{,i} + \chi^{(v)}_i$ with $\chi^{(v)i}_{\;\;\;\;\;\;,i} \equiv 0$. The MS takes $\kappa \equiv 0$, and the ZSG takes $\chi \equiv 0$ as the slicing condition; $\kappa$ is the trace of extrinsic curvature, and the tracefree part of extrinsic curvature vanishes for $\chi_i = 0$, see Equation (B7) in HNa.
In the MS, thus $\kappa \equiv 0$, this equation can be decomposed as
\bea
   & &
       \chi = - {12 \pi G \over c^3} \Delta^{-2} \nabla^i
       \left[ \left( \varrho + {p \over c^2} \right)
       \gamma^2 v_i + {1 \over c^2} \Pi_{ij} v^j \right],
   \label{mom-constraint-chi} \\
   & &
       \Delta \chi^{(v)}_i
       = - {16 \pi G \over c^3}
       \left\{ \left( \varrho + {p \over c^2} \right)
       \gamma^2 v_i + {1 \over c^2} \Pi_{ij} v^j
       - \nabla_i \Delta^{-1} \nabla^j \left[
       \left( \varrho + {p \over c^2} \right)
       \gamma^2 v_j + {1 \over c^2} \Pi_{jk} v^k \right] \right\}.
   \label{mom-constraint-chi-v}
\eea
Similarly, in the ZSG, thus $\chi \equiv 0$, we have
\bea
   & &
       \kappa = - {12 \pi G \over c^2} \Delta^{-1} \nabla^i
       \left[ \left( \varrho + {p \over c^2} \right)
       \gamma^2 v_i + {1 \over c^2} \Pi_{ij} v^j \right],
   \label{mom-constraint-kappa}
\eea
and $\chi^{(v)}_i$ is the same as in Equation (\ref{mom-constraint-chi-v}). From these we can estimate
\bea
   & &
       {1 \over c} \Delta^{-1} \kappa_{,i}
       \sim \chi_{,i} \sim \chi^{(v)}_i
       \sim \left( \gamma^2 {t_\ell^2 \over t_g^2} \right)
       {v_i \over c},
   \label{chi-estimate}
\eea
where it is understood that the MS and the ZSG consider $\kappa \equiv 0$ and $\chi \equiv 0$, respectively. Thus, under the action-at-a-distance condition $\Delta^{-1} \kappa_{,i}/c$, $\chi_{,i}$ and $\chi^{(v)}_i$ are negligible compared with $v_i/c$.

\subsection{Conservation equations}

Using Equations (\ref{assumptions}) and (\ref{chi-estimate}), the covariant momentum conservation in Equation (\ref{eq7-cov}) gives
\bea
   & &
       {1 \over \gamma} {d \over dt} \left( \gamma {\bf v} \right)
       = \nabla \Phi + {v^2 \over c^2} \nabla \Psi
       - {1 \over \varrho + p/c^2} \left( {1 \over \gamma^2} \nabla p + {\bf v} {1 \over c^2} {d {p} \over dt} \right)
   \nonumber \\
   & & \qquad
       + {1 \over \varrho + p/c^2} \left\{
       - {1 \over \gamma^2} \Pi^j_{i,j}
       + {1 \over c^2} \left[ {\bf v} \Pi_j^k v^j_{\;\;,k}
       - {1 \over \gamma^2} \left( \Pi_{ij} v^j \right)^{\displaystyle\cdot} \right]
       + {1 \over c^4} {\bf v} \Pi_{jk} v^j \dot v^k \right\},
   \label{M-conservation-1}
\eea
with ${\bf v} = v_i$. This can be arranged to give Equation (\ref{M-conservation}) and
\bea
   & &
       {d \over dt} \ln{\gamma}
       = {1 \over c^2} {\bf v} \cdot \nabla \Phi
       - {1 \over \varrho + p/c^2}
       {1 \over c^2}
       \left( {d p \over dt}
       - {1 \over \gamma^2} \dot {p} \right)
   \nonumber \\
   & & \qquad
       + {1 \over \varrho + p/c^2} {1 \over c^2} \left\{
       - {1 \over \gamma^2} v^i \Pi^j_{i,j}
       + {1 \over c^2} \left[ v^2 \Pi_j^k v^j_{\;\;,k}
       - {1 \over \gamma^2} v^i \left( \Pi_{ij} v^j \right)^{\displaystyle\cdot} \right]
       + {1 \over c^4} v^2 \Pi_{jk} v^j \dot v^k \right\}.
   \label{M-conservation-3}
\eea

The continuity equation $(\overline \varrho u^c)_{;c} = 0$ can be written as $\widetilde {\dot {\overline \varrho}} + \overline \varrho \theta = 0$, and leads to Equation (\ref{continuity}). We used
\bea
   & &
       \widetilde {\dot \varrho} \equiv c \varrho_{,c} u^c = \gamma {d \varrho \over dt}, \quad
       \theta \equiv c u^c_{\;\;;c} = \gamma \left[ \nabla \cdot {\bf v} + {d \over dt} \ln{\gamma} + {1 \over c^2} {\bf v} \cdot \nabla \left( \Psi - \Phi \right) \right],
\eea
which follow from Equations (C2), (C3), (D10) and (D11) in HNa. Equation (\ref{eq0}) also gives Equation (\ref{continuity}) directly.

Using Equation (\ref{M-conservation-3}) we have
\bea
   & &
       {d \overline \varrho \over dt}
       + \overline \varrho \nabla \cdot {\bf v}
       =
       - \overline \varrho {1 \over c^2}
       {\bf v} \cdot \nabla \Psi
       + {\overline \varrho \over c^2}
       {1 \over \varrho + p/c^2}
       \left( {d p \over dt}
       - {1 \over \gamma^2} \dot {p} \right)
   \nonumber \\
   & & \qquad
       - {1 \over c^2}
       {\overline \varrho \over \varrho + p/c^2} \left\{
       - {1 \over \gamma^2} v^i \Pi^j_{i,j}
       + {1 \over c^2} \left[ v^2 \Pi_j^k v^j_{\;\;,k}
       - {1 \over \gamma^2} v^i \left( \Pi_{ij} v^j \right)^{\displaystyle\cdot} \right]
       + {1 \over c^4}  \Pi_{jk} v^j \dot v^k \right\}.
   \label{Continuity-1}
\eea

The covariant energy conservation in Equation (\ref{eq6-cov}) gives Equation (\ref{E-conservation}). Using Equation (\ref{M-conservation-3}) we have
\bea
   & &
       {d \varrho \over dt}
       + \left( \varrho + {p \over c^2} \right)
       \nabla \cdot {\bf v}
       =
       - \left( \varrho + {p \over c^2} \right) {1 \over c^2}
       {\bf v} \cdot \nabla \Psi
       + {1 \over c^2} \left( {d p \over dt}
       - {1 \over \gamma^2} \dot {p} \right)
   \nonumber \\
   & & \qquad
       + {1 \over c^2} \left[ - 2 \Pi^j_i v^i_{\;,j}
       + {1 \over \gamma^2} \left( \Pi^j_i v^i \right)_{,j} \right]
       + {1 \over c^4} \left[ - 2 \Pi_{ij} v^i \dot v^j
       + {1 \over \gamma^2} \left( \Pi_{ij} v^i v^j
       \right)^{\displaystyle\cdot} \right].
   \label{E-conservation-1}
\eea

For later use, we present alternative expressions of the conservation equations
\bea
   & &
       {\partial \over \partial t} \left[ \left(
       \varrho + {p \over c^2} \right) \gamma^2 \right]
       + \nabla \cdot \left[ \left(
       \varrho + {p \over c^2} \right)
       \gamma^2 {\bf v} \right]
   \nonumber \\
   & & \qquad
       = \left( \varrho + {p \over c^2} \right)
       {1 \over c^2} \gamma^2 {\bf v} \cdot
       \nabla \left( 2 \Phi - \Psi \right)
       + {1 \over c^2} \dot {p}
       - {1 \over c^2} \nabla_i \left( \Pi^i_j v^j \right)
       - {1 \over c^4} \left( \Pi_{ij} v^i v^j \right)^{\displaystyle\cdot},
   \label{E-conservation-2} \\
   & &
       {\partial \over \partial t} \left[
       \left( \varrho + {p \over c^2} \right) \gamma^2 {\bf v} \right]
       + \nabla_j \left[
       \left( \varrho + {p \over c^2} \right) \gamma^2 {\bf v} v^j \right]
   \nonumber \\
   & & \qquad
       = \left( \varrho + {p \over c^2} \right) \gamma^2
       \left[ \nabla \Phi
       + {v^2 \over c^2} \nabla \Psi
       + {1 \over c^2} {\bf v} {\bf v} \cdot \nabla
       \left( \Phi - \Psi \right) \right]
       - \nabla p
       - \nabla_j \Pi^j_i
       - {1 \over c^2} \left( \Pi_{ij} v^j \right)^{\displaystyle\cdot},
   \label{M-conservation-4}
\eea
which also follow from Equations (\ref{eq6-ADM}) and (\ref{eq7-ADM}), respectively. These conservation equations are valid in both gauges.

The presence of the post-Newtonian potential $\Psi$ in Equations (\ref{continuity})-(\ref{M-conservation}) demands two Poisson's equations. In Hwang et al.\ (2016), however, we have argued that although the $(v^2/c^2)\nabla\Psi$ term in Equation (\ref{M-conservation-1}) is comparable to the $\nabla\Phi$ term in our approximation, it is negligible compared with the convective term ${\bf v} \cdot \nabla {\bf v}$ in the left-hand-side due to the weak gravity condition.
[In the ultra-relativistic case ($v \simeq c$) the $(v^2/c^2)\nabla\Phi$ term is, unless we have steep gradients in the gravitational potentials $\Phi$ and $\Psi$, negligible compared with the convective term by the weak gravity condition; in this case the $\nabla\Psi$ term can be negligible as well. In the relativistic case (we keep $v^2/c^2$ order, but $v^2 \ll c^2$) we have the $(v^2/c^2)\nabla\Psi$-term much smaller than the $\nabla \Phi$-term, and the latter is again much smaller than the convective term: thus although we may have to keep the $\nabla\Phi$ term, the $(v^2/c^2)\nabla\Psi$ term can be ignored.]
This applies similarly to $\Psi$ and $c^{-2} {\bf v} {\bf v} \cdot \nabla \Phi$ terms in Equations (\ref{continuity})-(\ref{Pi-equation}), (\ref{M-conservation-3}) and (\ref{E-conservation-1})-(\ref{M-conservation-4}), and $\nabla ( \Phi - \Psi)$ terms in Equations (\ref{E-conservation-2}) and (\ref{M-conservation-4}), see Equation (\ref{Phi-Psi}). 

Thus, the basic set of conservation equations effectively becomes Equations (\ref{continuity-practice})-(\ref{M-conservation-practice}). We have kept these in practice negligible terms because, first it is consistent to keep these terms in our derivation from Einstein's field equations, second these general equations might relevant in the presence of steep gradients of gravitational potentials, and third the results are also consistent with the 1PN expansion to be presented in the next section, see the last paragraph in Section \ref{sec:PN}.

Now we examine the remaining Einstein's field equations in the two gauge conditions separately, and show the full consistency of the derived equations.

\subsection{Einstein's equations in the MS}

In the MS, the ADM energy-constraint and the trace of ADM propagation in Equations (\ref{eq4}) and (\ref{eq2}) give Equations (\ref{Poisson-MS}) and (\ref{Poisson-Psi}), respectively. In order to show the consistency of these two equations we need to examine the tracefree ADM propagation in Equation (\ref{eq5}). This gives
\bea
   & &
       \left( \nabla^i \nabla_j - {1 \over 3} \delta^i_j \Delta \right)
       \left( \Phi - \Psi \right)
       - {8 \pi G \over c^2} \left[
       \left( \varrho + {p \over c^2} \right) \gamma^2 \left( v^i v_j
       - {1 \over 3} \delta^i_j v^2 \right)
       + \Pi^i_j - {1 \over 3} \delta^i_j \Pi^k_k \right]
   \nonumber \\
   & & \qquad
       = - c {\partial \over \partial t}
       \bigg[ \left( \nabla^i \nabla_j
       - {1 \over 3} \delta^i_j \Delta \right) \chi
       + {1 \over 2} \left( \chi^{(v)i}_{\;\;\;\;\;\;,j}
       + \chi^{(v),i}_j \right) \bigg]
   \nonumber \\
   & & \qquad
       = {4 \pi G \over c^2} \Delta^{-1} {\partial \over \partial t}
       \bigg\{ 2 \nabla_j \left[ \left( \varrho
       + {p \over c^2} \right) \gamma^2 v^i
       + {1 \over c^2} \Pi^i_k v^k \right]
       + 2 \nabla^i \left[ \left( \varrho
       + {p \over c^2} \right) \gamma^2 v_j
       + {1 \over c^2} \Pi_{jk} v^k \right]
   \nonumber \\
   & & \qquad \qquad
       - \left( \nabla^i \nabla_j \Delta^{-1} + \delta^i_j \right)
       \nabla_k \left[ \left( \varrho + {p \over c^2} \right)
       \gamma^2 v^k
       + {1 \over c^2} \Pi^k_\ell v^\ell \right] \bigg\}
   \nonumber \\
   & & \qquad
       = {4 \pi G \over c^2} \Delta^{-1} \bigg\{
       - 2 \nabla_j \nabla_k \left[ \left( \varrho
       + {p \over c^2} \right) \gamma^2 v^i v^k
       + \Pi^{ik} \right]
       - 2 \nabla^i \nabla_k \left[ \left( \varrho
       + {p \over c^2} \right) \gamma^2 v_j v^k
       + \Pi^k_j \right]
   \nonumber \\
   & & \qquad \qquad
       + \left( \nabla^i \nabla_j \Delta^{-1} + \delta^i_j \right)
       \nabla_k \nabla_\ell \left[
       \left( \varrho + {p \over c^2} \right) \gamma^2 v^k v^\ell
       + \Pi^{k \ell} \right]
       - 3 \left( \nabla^i \nabla_j
       - {1 \over 3} \delta^i_j \Delta \right) p
       \bigg\},
   \label{tracefree-ADM-propagation}
\eea
where we used Equations (\ref{mom-constraint-chi}), (\ref{mom-constraint-chi-v}) and (\ref{M-conservation-4}). Thus we have
\bea
   & &
       \Delta \left( \Phi - \Psi \right)
       = - {4 \pi G \over c^2} \left[ 3 p
       + \left( \varrho + {p \over c^2} \right)
       \gamma^2 v^2 + \Pi^i_i \right].
   \label{Phi-Psi}
\eea
Using Equation (\ref{Phi-Psi}), Equations (\ref{Poisson-MS}) and (\ref{Poisson-Psi}) become consistent. We note that $\chi_i$ has a nontrivial role only in Equations (\ref{tracefree-ADM-propagation}) and (\ref{Phi-Psi}); Equation (\ref{tracefree-ADM-propagation}) still has contribution from $\chi^{(v)}_i$ part, but the term has no role in deriving Equation (\ref{Phi-Psi}). In this sense, we have to keep $\chi^{(v)}_i$ in our metric although it has no role in deriving the final result.

One remaining equation to be checked for consistency is the trace of extrinsic curvature in Equation (\ref{eq1}). This gives
\bea
   & &
       c \Delta \chi = - 3 {\dot \Psi \over c^2}.
   \label{kappa-def}
\eea
We can show that this is consistent with Equations (\ref{Poisson-Psi}), (\ref{mom-constraint-chi}) and (\ref{E-conservation-2}). This completes the proof of full consistency of Equations (\ref{continuity})-(\ref{M-conservation}), (\ref{Poisson-MS}) and (\ref{Poisson-Psi}) in Einstein's gravity, under the assumptions of Equation (\ref{assumptions}).

\subsection{Einstein's equations in the ZSG}

The case of ZSG proceeds similarly. Equation (\ref{eq2}) gives Equation (\ref{Poisson-Psi}). Equation (\ref{eq2}) gives
\bea
   & &
       \Delta \Phi
       + 4 \pi G \left( \varrho + {3 p \over c^2} \right)
       = \dot \kappa
       - {8 \pi G \over c^2} \left[
       \left( \varrho + {p \over c^2} \right)
       \gamma^2 v^2 + \Pi^i_i \right].
\eea
From Equation (\ref{mom-constraint-kappa}), using Equation (\ref{M-conservation-4}), we can derive Equation (\ref{Poisson-ZSG}). In order to show the consistency between Equation (\ref{Poisson-Psi}) and Equation (\ref{Poisson-ZSG}) we need to examine Equation (\ref{eq5}). This gives
\bea
   & &
       \left( \nabla^i \nabla_j - {1 \over 3} \delta^i_j \Delta \right)
       \left( \Phi - \Psi \right)
       - {8 \pi G \over c^2} \left[
       \left( \varrho + {p \over c^2} \right) \gamma^2 \left( v^i v_j
       - {1 \over 3} \delta^i_j v^2 \right)
       + \Pi^i_j - {1 \over 3} \delta^i_j \Pi^k_k \right]
   \nonumber \\
   & & \qquad
       = - c {\partial \over \partial t}
       {1 \over 2} \left( \chi^{(v)i}_{\;\;\;\;\;\;,j}
       + \chi^{(v),i}_j \right)
   \nonumber \\
   & & \qquad
       = {8 \pi G \over c^2} \Delta^{-1} {\partial \over \partial t}
       \bigg\{ \nabla_j \left[ \left( \varrho
       + {p \over c^2} \right) \gamma^2 v^i
       + {1 \over c^2} \Pi^i_k v^k \right]
       + \nabla^i \left[ \left( \varrho
       + {p \over c^2} \right) \gamma^2 v_j
       + {1 \over c^2} \Pi_{jk} v^k \right]
   \nonumber \\
   & & \qquad \qquad
       - 2 \nabla^i \nabla_j \Delta^{-1}
       \nabla_k \left[ \left( \varrho + {p \over c^2} \right)
       \gamma^2 v^k
       + {1 \over c^2} \Pi^k_\ell v^\ell \right] \bigg\}
   \nonumber \\
   & & \qquad
       = {8 \pi G \over c^2} \Delta^{-1} \bigg\{
       - \nabla_j \nabla_k \left[ \left( \varrho
       + {p \over c^2} \right) \gamma^2 v^i v^k
       + \Pi^{ik} \right]
       - \nabla^i \nabla_k \left[ \left( \varrho
       + {p \over c^2} \right) \gamma^2 v_j v^k
       + \Pi^k_j \right]
   \nonumber \\
   & & \qquad \qquad
       + 2 \nabla^i \nabla_j \Delta^{-1}
       \nabla_k \nabla_\ell \left[
       \left( \varrho + {p \over c^2} \right) \gamma^2 v^k v^\ell
       + \Pi^{k \ell} \right]
       \bigg\},
   \label{tracefree-ADM-propagation-ZSG}
\eea
where we used Equations (\ref{mom-constraint-chi}), (\ref{mom-constraint-chi-v}) and (\ref{M-conservation-4}). Thus we have
\bea
   & &
       \Delta \left( \Phi - \Psi \right)
       = {12 \pi G \over c^2} \Delta^{-1} \nabla_i \nabla^j
       \left[ \left( \varrho + {p \over c^2} \right)
       \gamma^2 \left( v^i v_j - {1 \over 3} \delta^i_j v^2 \right)
       + \Pi^i_j - {1 \over 3} \delta^i_j \Pi^k_k \right].
   \label{Phi-Psi-ZSG}
\eea
Using Equation (\ref{Phi-Psi-ZSG}), we can show that Equations (\ref{Poisson-Psi}) and (\ref{Poisson-ZSG}) are consistent. We note that $\chi_i$ has a nontrivial role only in Equations (\ref{tracefree-ADM-propagation-ZSG}) and (\ref{Phi-Psi-ZSG}); Equation (\ref{tracefree-ADM-propagation-ZSG}) still has contribution from $\chi^{(v)}_i$ part, but the term has no role in deriving Equation (\ref{Phi-Psi}). In this sense, we have to keep $\chi^{(v)}_i$, thus $\chi_i$, in our metric although it has no role in deriving the final result.

One remaining equation to be checked for consistency is the trace of extrinsic curvature in Equation (\ref{eq1}). This gives
\bea
   & &
       \kappa = - 3 {\dot \Psi \over c^2}.
   \label{kappa-def-ZSG}
\eea
We can show that this is consistent with Equations (\ref{Poisson-Psi}), (\ref{mom-constraint-chi}) and (\ref{E-conservation-2}).

%
%
\section{1PN approximation: complete equations without fixing slicing}
                                           \label{sec:PN}

Now, we switch our subject to the hydrodynamic 1PN approximation. The main purpose of the presentation is to show the consistency of our special-relativity-with-gravity approximation presented in the previous sections, but the general hydrodynamic set of 1PN equations {\it without} fixing the hypersurface condition presented in this section is also new in the literature; previously it was derived in the cosmological context (HNP). The complete set of 1PN equations without taking the hypersurface condition is
\bea
   & &
       \dot {\overline \varrho} + \nabla \cdot
       \left( \overline \varrho \overline {\bf v} \right)
       = - {1 \over c^2} \overline \varrho
       \left( {\partial \over \partial t}
       + \overline {\bf v} \cdot \nabla \right)
       \left( {1 \over 2} \overline v^2 + 3 U \right),
   \label{1PN-0} \\
   & &
       \dot {\overline \varrho} + \nabla \cdot
       \left( \overline \varrho \overline {\bf v} \right)
       = - {1 \over c^2} \left[ \overline \varrho
       \left( {\partial \over \partial t}
       + \overline {\bf v} \cdot \nabla \right)
       \left( {1 \over 2} \overline v^2 + 3 U + \Pi \right)
       + p \nabla \cdot \overline {\bf v}
       + \Pi_i^j \nabla_j v^i \right],
   \label{1PN-1} \\
   & &
       \dot {\overline {\bf v}}
       + \overline {\bf v} \cdot \nabla \overline {\bf v}
       - \nabla U + {1 \over \overline \varrho}
       \left( \nabla p
       + \nabla_j \Pi_i^j \right)
   \nonumber \\
   & & \qquad
       = {1 \over c^2} \bigg\{
       - 2 \nabla \left( U^2 - \widetilde \Phi \right)
       + \dot P_i + \overline v^j \left( P_{i,j} - P_{j,i} \right)
       - \overline {\bf v}
       \left( {\partial \over \partial t}
       + \overline {\bf v} \cdot \nabla \right)
       \left( {1 \over 2} \overline v^2 + 3 U \right)
       + \overline v^2 \nabla U
   \nonumber \\
   & & \qquad
       + {1 \over \overline \varrho}
       \left[
       \left( \overline v^2 + 4 U + \Pi + {p \over \overline \varrho} \right) \left( \nabla p + \nabla_j \Pi_i^j \right)
       - \overline {\bf v}
       \left( {\partial \over \partial t}
       + \overline {\bf v} \cdot \nabla \right) p
       - \left( \Pi_i^j \overline v^j \right)^{\displaystyle\cdot}
       - \overline {\bf v} \Pi_k^j \nabla_j v^k
       + 2 U \nabla_j \Pi_i^j
       \right] \bigg\},
   \label{1PN-2} \\
   & &
       \Delta U + 4 \pi G \overline \varrho
       = - {1 \over c^2} \bigg[ 3 \ddot U
       - 2 U \Delta U
       + 2 \Delta \widetilde \Phi
       + \dot P^i_{\;\;,i}
       + 8 \pi G \left( \overline \varrho \overline v^2 + {1 \over 2} \overline \varrho \Pi
       + {3 \over 2} p \right) \bigg],
   \label{1PN-3} \\
   & &
       0 = {1 \over 4} \left( P^j_{\;\;,ji} - \Delta P_i \right)
         + \nabla \dot U - 4 \pi G \overline \varrho \overline {\bf v},
   \label{1PN-4} \\
   & &
       0 = U - V.
   \label{1PN-5}
\eea
These follow from the continuity, the energy conservation, the momentum conservation, the trace of ADM propagation, the ADM momentum constraint, and the tracefree ADM propagation equations, respectively: see Equations (62), (114), (115), (119), (120) and (91), respectively, in HNP which were presented in the cosmological background\footnote{There are typos in the right-hand-side of Equation (115) of the paper. It should be read as
\bea
   & &
       = - {1 \over a \sigma} \Pi^j_{i|j}
       - {1 \over c^2} \dots
   \nonumber
\eea
};
in our non-cosmological background, we set $a \equiv 1$ and $\varrho_b \equiv 0 \equiv p_b$, etc. for the background quantities. As mentioned, these 1PN equations also follow from the fully nonlinear perturbation formulation (Noh \& Hwang 2013).

The left-hand and right-hand sides of Equations (\ref{1PN-0})-(\ref{1PN-5}) are the 0PN (Newtonian) and 1PN orders, respectively. From Equations (\ref{1PN-0}) and (\ref{1PN-1}) we have ${d\Pi/dt} + (p/\overline \varrho) \nabla \cdot {\bf v} = 0$. Our 1PN metric convention follows Chandrasekhar (1965)
\bea
   & &
       g_{00} = - \left[ 1 - {1 \over c^2} 2 U
       + {1 \over c^4} \left( 2 U^2
       - 4 \widetilde \Phi \right) \right], \quad
       g_{0i} = - {1 \over c^3} P_i, \quad
       g_{ij} = \left( 1 + {1 \over c^2} 2 V \right) \delta_{ij},
   \label{metric-1PN}
\eea
where the index of $P_i$ is raised and lowered by $\delta_{ij}$ as the metric. Compared with Equation (\ref{metric}) we have
\bea
   & &
       \Phi = U
       - {1 \over c^2} \left( U^2 - 2 \widetilde \Phi \right), \quad
       \chi_i = {1 \over c^3} P_i, \quad
       \Psi = V.
\eea
The velocity in the 1PN equations is defined as
\bea
   & &
       u^i \equiv u^0 {\overline v^i \over c},
\eea
where $\overline v^i$ is the fluid coordinate three-velocity (HNa); the index of $\overline v_i$ is raised and lowered by $\delta_{ij}$ as the metric. Compared with $v_i$ defined in Equation (\ref{v_i}) we have
\bea
   & &
       v_i = \overline v_i
       + {1 \over c^2} \left[ \left( U + 2 V \right) \overline v_i
       - P_i \right].
\eea

The above set of equations is still in a gauge-ready form, i.e., without taking the temporal gauge condition yet. The general 1PN hypersurface condition is (HNP)
\bea
   & &
       P^i_{\;\;,i} + n \dot U = 0.
   \label{1PN-gauge}
\eea
The choice of $n$ corresponds to taking different temporal gauges: notable ones are the harmonic gauge ($n \equiv 4$) in Weinberg (1972), the MS ($n \equiv 3$) in Chandrasekhar (1965) and the ZSG ($n \equiv 0$) (HNP).

Now, we show the gauge dependence of $3p/c^2$ term in the Poisson's equation (\ref{1PN-3}). To 0PN order Equation (\ref{1PN-3}) gives $\Delta U = - 4 \pi G \overline \varrho$. From this, using Equation (\ref{1PN-1}), we have $\Delta \dot U = 4 \pi G \nabla \cdot \left( \overline \varrho \overline {\bf v} \right)$ which also follows from Equation (\ref{1PN-4}). From this, using Equations (\ref{1PN-1}) and (\ref{1PN-2}), we can show
\bea
   & &
       \ddot U
       = - 4 \pi G p
       + 4 \pi G \Delta^{-1} \left[ \nabla \cdot
       \left( \overline \varrho \nabla U \right)
       - \nabla_i \nabla_j
       \left( \overline \varrho \overline v^i \overline v^j
       + \Pi^{ij}
       \right) \right].
   \label{ddot-U}
\eea
Imposing the general gauge condition in Equation (\ref{1PN-gauge}), thus still in a gauge-ready form, Equation (\ref{1PN-3}) gives
\bea
   & &
       \Delta U + 4 \pi G \varrho
       + {1 \over c^2} \big[
       - (n - 3) \ddot U + 12 \pi G p + \dots \big] = 0.
   \label{PN-Poisson-2}
\eea
Using Equation (\ref{ddot-U}), Equation (\ref{PN-Poisson-2}) can be arranged as
\bea
   & &
       \Delta U + 4 \pi G \varrho + {1 \over c^2} \big[
       4 \pi G n p + \dots \big]  = 0.
   \label{PN-Poisson-3}
\eea
From Equations (\ref{PN-Poisson-2}) and (\ref{PN-Poisson-3}) we notice that both the propagation speed of the gravitational potential and the presence of a pressure term in the 1PN Poisson's equation depend on the gauge condition, and these are related. Under the general gauge condition the propagation speed can be na\"ively read from Equation (\ref{PN-Poisson-2}) as $c/\sqrt{n-3}$ which becomes the speed of light only in the harmonic gauge with $n = 4$, and the pressure term can be read from Equation (\ref{PN-Poisson-3}) as $n p/c^2$.

The gauge dependence of the propagation speed of gravitational potential is not surprising because the potential is gauge dependent. Whereas the propagation speed of the (physical) Weyl tensor naturally does not depend on the gauge choice and is always $c$: see Section 7 of HNP. We have an exact analogy in the classical electromagnetism where the propagation speed of the field potential depends on the gauge choice, whereas the propagation speed of the (physical) field strength is always $c$ (Jackson 2002).

Likewise, the form of 1PN pressure term also depends on the gauge choice. It vanishes in the ZSG, and becomes $3 p/c^2$ and $4 p/c^2$ in the MS and the harmonic gauge, respectively. In the MS with $P^i_{\;\;,i} = - 3 \dot U$ the $3 \ddot U$ in Equation (\ref{1PN-3}) cancels by the $\dot P^i_{\;\;,i}$, thus leaving $12 \pi G p$. Similarly, in the ZSG with $ P^i_{\;\;,i} = 0$, the $12 \pi G p$ in Equation (\ref{1PN-3}) cancels by the $3 \ddot U$ using Equation (\ref{ddot-U}).

We can show that Equations (\ref{continuity})-(\ref{M-conservation}), (\ref{Poisson-MS}), (\ref{Poisson-Psi}) and (\ref{Poisson-ZSG}) in the two gauge conditions are {\it consistent} with the 1PN equations (\ref{1PN-0})-(\ref{1PN-5}), in the overlapping regime (i.e., for the weak gravity and the action-at-a-distance, and 1PN for the pressure, anisotropic stress and velocity) of the two approximations. In both gauges, according to the comment below Equation (\ref{assumptions}) we have ${\bf v} = \overline {\bf v}$, but $\overline {\bf v} \cdot \nabla \overline {\bf v} = {\bf v} \cdot \nabla {\bf v} - 3 {\bf v} {\bf v} \cdot \nabla \Phi/c^2$ and $\nabla \cdot (\overline \varrho \overline {\bf v}) = \nabla \cdot (\overline \varrho {\bf v}) - 3 \overline \varrho {\bf v} \cdot \nabla \Phi/c^2$. In the MS the Equations (\ref{1PN-0})-(\ref{1PN-5}) become
\bea
   & & \dot {\overline \varrho}
       + \nabla \cdot \left( {\overline \varrho} {\bf v} \right)
       = {1 \over c^2} \left(
       - {\overline \varrho} {\bf v} \cdot \nabla \Phi
       + {\bf v} \cdot \nabla p
       + v^i \nabla_j \Pi^j_i \right),
   \label{1PN-continuity} \\
   & & \dot {\overline \varrho}
       + \nabla \cdot \left( {\overline \varrho} {\bf v} \right)
       = {1 \over c^2} \left(
       - {\overline \varrho} {\bf v} \cdot \nabla \Phi
       - {\overline \varrho} {d \Pi \over dt}
       + {\bf v} \cdot \nabla p
       - p \nabla \cdot {\bf v}
       + v^i \nabla_j \Pi^j_i
       - \Pi^j_i \nabla_j v^i \right),
   \label{1PN-E-conservation} \\
   & & {d \over dt} {\bf v}
       - \nabla \Phi
       + {1 \over {\overline \varrho}} \left(
       \nabla p + \nabla_j \Pi_i^j \right)
   \nonumber \\
   & & \qquad
       = {1 \over c^2} \left\{
       v^2 \nabla \Phi
       - {\bf v} {\bf v} \cdot \nabla \Phi
       + {1 \over {\overline \varrho}}
       \left[
       - {\bf v} \dot p
       + \left( v^2 + \Pi + {p \over \overline \varrho} \right)
       \left( \nabla p
       + \nabla_j \Pi_i^j \right)
       + {\bf v} \nabla_j \left( \Pi^j_k v^k \right)
       - \left( \Pi_i^j v_j \right)^{\displaystyle\cdot}
       \right] \right\},
   \label{1PN-M-conservation} \\
   & & \Delta \Phi
       + 4 \pi G \left( \varrho + 3 {p \over c^2} \right)
       = - {8 \pi G \over c^2} \overline \varrho v^2,
   \label{1PN-Poisson-MS} \\
   & & 0
       = \Delta \Psi
       + 4 \pi G \overline \varrho,
   \label{1PN-Poisson-Psi} \\
   & & 0
       = \nabla_i \dot \Phi
       - 4 \pi G \overline \varrho v_i
       - {1 \over 4} c^3 \Delta \chi_i^{(v)}.
   \label{1PN-Phi-dot}
\eea
Equation (\ref{1PN-Phi-dot}) is consistent with Equations (\ref{kappa-def}), (\ref{kappa-def-ZSG}), (\ref{mom-constraint-chi}) and (\ref{mom-constraint-kappa}) valid to the 0PN order. In the ZSG instead of Equation (\ref{1PN-Poisson-MS}) we have
\bea
   \Delta \Phi + 4 \pi G \varrho
       = - {8 \pi G \over c^2} \overline \varrho v^2
       + {12 \pi G \over c^2}
       \Delta^{-1} \nabla_i \nabla_j \left( \overline \varrho v^i v^j
       + \Pi^{ij} \right),
   \label{1PN-Poisson-ZSG}
\eea
and the other equations remain the same.

%
%
\section{Gauge transformation}
                                         \label{sec:GT}

Now, we can show that to the linear order in perturbation the presence or absence of $3p/c^2$ term in the extended Newtonian Poisson's equation in Equations (\ref{Poisson-MS}) or (\ref{Poisson-ZSG}) depends on the temporal gauge while Equations (\ref{continuity-practice})-(\ref{M-conservation-practice}) and (\ref{Poisson-Psi}) remain the same. Under the gauge transformation $\widehat x^c = x^c + \xi^t$, in a non-expanding background, the gauge transformation properties in Equation (252) of Noh \& Hwang (2004) give
\bea
   & &
       \widehat \Phi = \Phi + c \dot \xi^t, \quad
       \widehat \Psi = \Psi, \quad
       \widehat \chi = \chi - \xi^t, \quad
       \widehat \kappa = \kappa + c \Delta \xi^t, \quad
       \widehat \varrho = \varrho, \quad
       \widehat p = p, \quad
       \widehat v = v - c \xi^t,
   \label{GT-1}
\eea
where the spatial gauge transformation is already fixed by our spatial gauge condition (Bardeen 1988); similarly we can add $\widehat {\overline \varrho} = \overline \varrho$ for the rest mass density; we introduced $v_i \equiv v_{,i} + v_i^{(v)}$ with $v^{(v)i}_{\;\;\;\;\;\;,i} \equiv 0$. Thus, variables $\Psi$, $\overline \varrho$, $\varrho$ and $p$ are invariant under the gauge transformation. Lets consider $\widehat x^c$ and $x^c$ coordinates as the MS and the ZSG, respectively. Then using $\chi_{\rm ZSG} \equiv 0$ and $\kappa_{\rm MS} \equiv 0$, we have
\bea
   & &
       \Phi_{\rm MS} = \Phi_{\rm ZSG} + c \dot \xi^t, \quad
       \chi_{\rm MS} = - \xi^t, \quad
       \kappa_{\rm ZSG} = - c \Delta \xi^t, \quad
       v_{\rm MS} = v_{\rm ZSG} - c \xi^t,
   \label{GT-2}
\eea
and $\Psi_{\rm MS} = \Psi_{\rm ZSG}$ etc. Using these gauge transformation properties we can show the following. To the linear order Equation (\ref{Poisson-Psi}) is apparently gauge invariant. The forms of Equations (\ref{continuity-practice})-(\ref{M-conservation-practice}) to the linear order remain the same in both gauges in our approximation in Equation (\ref{assumptions}). From Equation (\ref{GT-2}) we have
\bea
   & &
       \Delta \Phi_{\rm MS} - \Delta \Phi_{\rm ZSG}
       = c \Delta \dot \xi^t
       = - c \Delta \dot \chi_{\rm MS}
       = - \dot \kappa_{\rm ZSG}
       = 3 {\ddot \Psi \over c^2}
       = - 12 \pi G {p \over c^2},
   \label{GT-Phi}
\eea
where we used Equations (\ref{kappa-def}), (\ref{kappa-def-ZSG}), (\ref{E-conservation-practice})-(\ref{Poisson-MS}) and (\ref{Poisson-ZSG}), and the approximation in Equation (\ref{assumptions}). Thus, we exactly recover the pressure difference.

%
%
%
\section{Another expressions}

In a conventional notation of special relativistic hydrodynamics suitable for numerical study (Wilson \& Mathews 2003; Ryu, Chattopadhyay \& Choi 2006), Equations (\ref{continuity})-(\ref{M-conservation}) can be arranged using Equations (\ref{E-conservation-2}) and (\ref{M-conservation-4}), as
\bea
   & &
       \dot D + \nabla_i \left( D v^i \right)
       = {1 \over c^2} D {\bf v} \cdot \nabla
       \left( \Phi - \Psi \right),
   \label{continuity-SR} \\
   & &
       \dot E
       + \nabla \cdot \left[ \left( E
       + p \right) {\bf v} \right]
       = {1 \over c^2} \left( E + p \right) {\bf v} \cdot
       \left[ \nabla \Phi
       + \nabla \left( \Phi - \Psi \right)
       + {v^2 \over c^2} \nabla \Psi \right]
       - \nabla_j \left( \Pi_i^j v^i \right)
       - {1 \over c^2} \left( \Pi_{ij} v^i v^j \right)^{\displaystyle\cdot},
   \label{E-conservation-SR} \\
   & &
       \dot M_i + \nabla_j \left( M_i v^j
       + p \delta^j_i \right)
       = {1 \over c^2} \left( E + p \right)
       \left[ \nabla_i \Phi
       + {1 \over c^2} v_i {\bf v} \cdot \nabla
       \left( \Phi - \Psi \right)
       + {v^2 \over c^2} \nabla_i \Psi \right]
       - \nabla_j \Pi_i^j
       - {1 \over c^2} \left( \Pi_{ij} v^j \right)^{\displaystyle\cdot},
   \label{M-conservation-SR}
\eea
where
\bea
   & &
       D \equiv \gamma \overline \varrho, \quad
       E \equiv \gamma^2 \overline \varrho c^2 h - p, \quad
       M_i \equiv \gamma^2 \overline \varrho h v_i,
\eea
are the coordinate (Eulerian) mass density, the ADM energy density and the coordinate momentum density, respectively, with $h \equiv 1 + \Pi/c^2 + p/(\overline \varrho c^2)$ the specific enthalpy; we have $E + p = \gamma^2 (\varrho c^2 + p)$. The gravitational potential $\Phi$ is determined by Equation (\ref{Poisson-MS}) in the MS and Equation (\ref{Poisson-ZSG}) in the ZSG, and $\Psi$ by Equation (\ref{Poisson-Psi}). In terms of the above notation we have
\bea
   & &
       \Delta \Phi
       + 4 \pi G \left[ {1 \over c^2} \left( E + 3 p \right)
       + {M^i M_i \over E + p} + {2 \over c^2} \Pi^i_i \right]
       = 0,
   \label{Poisson-MS-E} \\
   & &
       \Delta \Psi
       + {4 \pi G \over c^2} \left( E+ \Pi^i_i \right)
       = 0,
   \label{Poisson-Psi-E}
\eea
in the MS; in the case of ZSG, instead of Equation (\ref{Poisson-MS-E}) we have
\bea
   & &
       \Delta \Phi
       + 4 \pi G \left[
       {1 \over c^2} E
       + {M^i M_i \over E + p} + {2 \over c^2} \Pi^i_i
       - 3 \Delta^{-1} \nabla_i \nabla_j
       \left( {M^i M^j \over E + p} + \Pi^{ij} \right) \right]
       = 0.
   \label{Poisson-ZSG-E}
\eea

If we strictly apply the weak gravity limit in Equation (\ref{assumptions}), we can ignore the $\nabla (\Phi - \Psi)$ and $\nabla \Psi$ terms in Equations (\ref{continuity-SR})-(\ref{M-conservation-SR}). Although these terms are consistently derived in our special-relativity-with-gravity approximation, as we argued below Equation (\ref{M-conservation-4}), the strengths of these terms are in practice negligible compared with other terms in the equations. Thus, in practice, we have
\bea
   & &
       \dot D + \nabla_i \left( D v^i \right)
       = 0,
   \label{continuity-SR-practice} \\
   & &
       \dot E
       + \nabla_i \left[ \left( E
       + p \right) v^i \right]
       = {1 \over c^2} \left( E + p \right) {\bf v} \cdot
       \nabla \Phi
       - \nabla_j \left( \Pi_i^j v^i \right)
       - {1 \over c^2} \left( \Pi_{ij} v^i v^j \right)^{\displaystyle\cdot},
   \label{E-conservation-SR-practice} \\
   & &
       \dot M_i + \nabla_j \left( M_i v^j
       + p \delta^j_i \right)
       = {1 \over c^2} \left( E + p \right)
       \nabla_i \Phi
       - \nabla_j \Pi_i^j
       - {1 \over c^2} \left( \Pi_{ij} v^j \right)^{\displaystyle\cdot},
   \label{M-conservation-SR-practice}
\eea
which correspond to Equations (\ref{continuity-practice})-(\ref{M-conservation-practice}); $\Phi$ is determined by Equation (\ref{Poisson-MS-E}) in the MS, and by Equation (\ref{Poisson-ZSG-E}) in the ZSG. The gravity term in Equation (\ref{E-conservation-SR-practice}) is present even in Newtonian hydrodynamics, see Equation (2.32) in Shu (1992); this is because $E$ contains both $\overline \varrho c^2$ order term together with $\overline \varrho \Pi$ and $p$ terms\footnote{We thank professor Dongsu Ryu for pointing out this.
                       },
thus in order to derive this term we have to use Equation (\ref{continuity}) instead of Equation (\ref{continuity-practice}).

%
%
\section{Discussion}
                                             \label{sec:Discussion}

We presented two hydrodynamic approximations of Einstein's gravity. The first method, which is new, is the special relativistic hydrodynamics with fully relativistic pressure and velocity in the presence of self-gravity. In the MS Equations (\ref{continuity-practice})-(\ref{Poisson-MS}) are the complete set, and in the ZSG Equation (\ref{Poisson-ZSG}) replaces Equation (\ref{Poisson-MS}).

The second method is the 1PN approximation: our new contribution here is the gauge-ready expression in a non-expanding background.
Previously, the hydrodynamic 1PN approximation was studied by Chandrasekhar (1965) in the MS and by Weinberg (1972) in the harmonic gauge; the gauge-ready extension was made by us in the cosmological context (HNP). Equations (\ref{1PN-0})-(\ref{1PN-5}) are the complete set of 1PN equations with the general temporal-gauge condition in Equation (\ref{1PN-gauge}).

The fully general relativistic hydrodynamic equations in terms of $D$, $E$ and $M_i$ are known in the literature (Wilson \& Mathews 2003; Alcubierre 2008; Bona, Palenzuela-Luque \& Bona-Casas 2009; Baumgarte \& Shapiro 2010,; Gourgoulhon 2012; Rezzolla \& Zanotti 2013). The numerical relativity generally demands heavy computational resources with some conceptual difficulties compared with the Newtonian hydrodynamics. Our special relativistic hydrodynamics with self-gravity is formally quite similar to the Newtonian one. The special relativistic hydrodynamics with weak gravity was not known in the literature.

We can imagine many astrophysical situations where the special relativistic effect is important while the gravity is weak. Except near neutron stars or black holes, in most celestial bodies (planets, stars, star cluster, galaxies, galaxy clusters, interstellar and intergalactic media, and even including astrophysical regions some distances away from pulsar, quasars, jets, etc) the gravity is weak. The strength of the gravity is measured by $\Phi/c^2 \sim GM/(\ell c^2)$ with $M$ and $\ell$ the characteristic mass and size dimensions, respectively. In the celestial bodies mentioned above this dimensionless gravitational strength is of the order $\Phi/c^2 \sim 10^{-5} - 10^{-10}$ and even smaller. In such situations the gravity is indeed weak. We anticipate our two approximations may find wide applications in such astrophysical situations where the special relativistic activity is important.

In most realistic astrophysical situation where special relativity is important, however, often the magnetic activity is also involved. Extending our special-relativity-with-gravity approximation including special relativistic MHD (magnetohydrodynamics) is currently in progress.

%
%
\acknowledgments

We would like to thank the anonymous referee for clarifying and insightful comments which helped to much improve the manuscript. We thank professor Dongsu Ryu for useful suggestions and critical comments on special relativistic hydrodynamics. J.H.\ was supported by Basic Science Research Program through the National Research Foundation (NRF) of Korea funded by the Ministry of Science, ICT and future Planning (No.\ 2013R1A2A2A01068519 and No.\ 2016R1A2B4007964). H.N.\ was supported by National Research Foundation of Korea funded by the Korean Government (No.\ 2015R1A2A2A01002791).

\appendix
%
%
\section{Appendix A. Special Relativistic Hydrodynamics}

The special relativistic hydrodynamic equations without the anisotropic stress are known in the literature (Weinberg 1972; Battaner 1996). Here we present the full equations in the presence of the anisotropic stress. We consider Minkowsky spacetime, thus $g_{ab} = \eta_{ab}$. The general energy-momentum tensor is in Equation (\ref{Tab}). The normalized ($u^c u_c \equiv -1$) four-vector is
\bea
       u_i \equiv \gamma {v_i \over c}, \quad
       u^i = \gamma {v^i \over c}, \quad
       u_0 = - u^0 = - \gamma; \quad
       \gamma \equiv - n_c u^c
       = {1 \over \sqrt{ 1 - {v^2 \over c^2}}},
\eea
with the index of $v^i$ raised and lowered by $\delta_{ij}$ and its inverse; we have $v^2 \equiv |{\bf v}|^2 \equiv v^i v_i$ with ${\bf v} \equiv v^i$; the normal four-vector is
\bea
       n_i \equiv 0 = n^i, \quad
       n_0 = - n^0 = -1.
\eea
The anisotropic stress ($\pi_{ab} u^b \equiv 0$) gives
\bea
   & &
       \pi_{ij} \equiv \Pi_{ij}, \quad
       \pi_{0i} = - \Pi_{ij} {v^j \over c}, \quad
       \pi_{00} = \Pi_{ij} {v^i v^j \over c^2}; \quad
       \pi^i_j = \Pi^i_j, \quad
       \pi^0_i = \Pi_{ij} {v^j \over c}, \quad
       \pi^i_0 = - \Pi^i_j {v^j \over c}, \quad
       \pi^0_0 = - \Pi_{ij} {v^i v^j \over c^2},
\eea
with the index of $\Pi_{ij}$ raised and lowered by $\delta_{ij}$ and its inverse. The condition $\pi^c_c \equiv 0$ gives
\bea
      \Pi^i_i = \Pi_{ij} {v^i v^j \over c^2}.
\eea

The continuity equation $(\overline \varrho u^c)_{;c} \equiv 0$ gives
\bea
       \left( \overline \varrho \gamma \right)^{\displaystyle\cdot}
       + \nabla \cdot \left( \overline \varrho \gamma {\bf v} \right) = 0.
\eea
The energy and momentum conservation equation $T^b_{a;b} \equiv 0$, for $a=0$ and $i$, respectively, give
\bea
   & &
       \left[ \left( \varrho + {p \over c^2} \right) \gamma^2
       \right]^{\displaystyle\cdot}
       + \nabla \cdot \left[ \left( \varrho + {p \over c^2} \right)
       \gamma^2 {\bf v} \right]
       = {\dot p \over c^2}
       - {1 \over c^2} \nabla_i \left( \Pi^i_j v^j \right)
       - {1 \over c^4} \left( \Pi_{ij} v^i v^j \right)^{\displaystyle\cdot},
   \label{SR-1} \\
   & &
       \left[ \left( \varrho + {p \over c^2} \right) \gamma^2 v_i
       \right]^{\displaystyle\cdot}
       + \nabla \cdot \left[ \left( \varrho + {p \over c^2} \right)
       \gamma^2 v_i {\bf v} \right]
       = - \nabla_i p
       - \Pi^j_{i,j}
       - {1 \over c^2} \left( \Pi_{ij} v^j \right)^{\displaystyle\cdot}.
   \label{SR-2}
\eea
From these we can derive alternative forms
\bea
   & &
       \left( \varrho + {p \over c^2} \right) \gamma^2
       {d v_i \over dt}
       = - \nabla_i p
       - v_i {\dot p \over c^2}
       - \Pi^j_{i,j}
       + {1 \over c^2} v_i \nabla_j \left( \Pi^j_k v^k \right)
       - {1 \over c^2} \left( \Pi_{ij} v^j \right)^{\displaystyle\cdot}
       + {1 \over c^4} v_i \left( \Pi_{jk} v^j v^k \right)^{\displaystyle\cdot},
   \label{SR-3} \\
   & &
       \left( \varrho + {p \over c^2} \right)
       {d \over dt} \ln{\gamma}
       = - {1 \over c^2} \left( {d p \over dt}
       - {1 \over \gamma^2} \dot p \right)
       - {1 \over c^2} \left[ v^i \Pi^j_{i,j}
       - {v^2 \over c^2}
       \left( \Pi^j_i v^i \right)_{,j} \right]
       - {1 \over c^4} \left[ v^i \left( \Pi_{ij} v^j \right)^{\displaystyle\cdot}
       - {v^2 \over c^2}
       \left( \Pi_{ij} v^i v^j \right)^{\displaystyle\cdot}
       \right],
   \label{SR-4} \\
   & &
       \left( \varrho + {p \over c^2} \right) {1 \over \gamma}
       {d \over dt} \left( \gamma v_i \right)
       = - \left( {1 \over \gamma^2} \nabla_i p
       + {v_i \over c^2} {d p \over dt} \right)
       - {1 \over \gamma^2} \left[ \Pi^j_{i,j}
       + {1 \over c^2} \left( \Pi_{ij} v^j \right)^{\displaystyle\cdot}
       \right]
       + {1 \over c^2} v_i \Pi_j^k
       \left( v^j_{\;,k}
       + {1 \over c^2} v^j \dot v_k \right),
   \label{SR-5} \\
   & &
       {d \varrho \over dt}
       + \left( \varrho + {p \over c^2} \right)
       \left( \nabla \cdot {\bf v}
       + {d \over dt} \ln{\gamma} \right)
       = - {1 \over c^2} \Pi^j_i v^i_{\;,j}
       - {1 \over c^4} \Pi_{ij} v^i \dot v^j.
   \label{SR-7} \\
   & &
       \overline \varrho {d \Pi \over dt}
       + p \left( \nabla \cdot {\bf v}
       + {d \over dt} \ln{\gamma} \right)
       = - \Pi^j_i v^i_{\;,j}
       - {1 \over c^2} \Pi_{ij} v^i \dot v^j.
   \label{SR-8}
\eea
The covariant energy and momentum conservation equations in Equations (\ref{eq6-cov}) and (\ref{eq7-cov}) give Equations (\ref{SR-7}) and (\ref{SR-5}), respectively. The ADM energy and momentum conservation equations in Equations (\ref{eq6-ADM}) and (\ref{eq7-ADM}) give Equations (\ref{SR-1}) and (\ref{SR-2}), respectively.

%
%
\section{Appendix B. Fully nonlinear perturbations in non-expanding background}

Here, we present the fully nonlinear and exact perturbation equations in the Minkowsky background {\it ignoring} the transverse-tracefree tensor-type perturbation. As the metric convention we take
\bea
   & &
       ds^2 = - \left( 1 + 2 \alpha \right) c^2 d t^2
       - 2 \chi_i c dt d x^i
       + \left( 1 + 2 \varphi \right) \delta_{ij} d x^i d x^j,
   \label{metric-FNL}
\eea
where $\alpha$, $\varphi$ and $\chi_i$ are functions of spacetime with arbitrary amplitude; the index of $\chi_i$ is raised and lowered by $\delta_{ij}$ as the metric. Without losing any generality, we have taken the spatial gauge condition as in Equation (\ref{spatial-gauge}). We can derive the exact inverse metric and the rest of the algebra is straightforward (HNa, Noh 2015). The lapse function and the Lorentz factor are
\bea
   & &
       {\cal N} \equiv {1 \over \sqrt{-g^{00}}}
       = \sqrt{ 1 + 2 \alpha
       + {\chi^k \chi_k \over 1 + 2 \varphi}}, \quad
       \gamma \equiv - u^c n_c = {1 \over \sqrt{ 1 - { v^k v_k
       \over c^2 (1 + 2 \varphi)}}},
\eea
where $v^i$ is introduced as in Equation (\ref{v_i}). The equations follow from the cosmological ones presented in Equations (6)-(11) and (14)-(19) of Hwang, Noh \& Park (2016); compared with our previous notations in the cosmological context we now set $a \equiv 1$ and $\varrho = \widehat \varrho$, $p = \widehat p$, $v_i = \widehat v_i$, etc. We have not taken the temporal gauge condition yet; as the temporal gauge condition we may impose the synchronous gauge ($\alpha \equiv 0$), the zero-shear gauge ($\chi^i_{\;\;,i} \equiv 0$), the comoving gauge ($v^i_{\;\;,i} \equiv 0$), and the uniform-expansion gauge or the maximal slicing ($\kappa \equiv 0$), see Equation (\ref{GT-1}); the synchronous gauge leaves remnant gauge mode even after imposing the gauge condition, see Equation (\ref{GT-1}).

\noindent
Definition of $\kappa$ (the trace of extrinsic curvature, $K^i_i$): \bea
   & & \kappa
       \equiv
       - {1 \over {\cal N} (1 + 2 \varphi)}
       \left[ 3 \dot \varphi
       + c \left( \chi^k_{\;\;,k}
       + {\chi^{k} \varphi_{,k} \over 1 + 2 \varphi} \right)
       \right].
   \label{eq1}
\eea
ADM energy constraint:
\bea
   & & {4 \pi G \over c^2} \mu
       + {c^2 \Delta \varphi \over (1 + 2 \varphi)^2}
       = {1 \over 6} \kappa^2
       - {4 \pi G \over c^2} \left( \mu + p \right)
       \left( \gamma^2 - 1 \right)
       + {3 \over 2} {c^2 \varphi^{,i} \varphi_{,i} \over (1 + 2 \varphi)^3}
       - {c^2 \over 4} \overline{K}^i_j \overline{K}^j_i
       - {1 \over (1 + 2 \varphi)^2}
       {4 \pi G \over c^4} \Pi_{ij}
       v^i v^j.
   \label{eq2}
\eea
ADM momentum constraint:
\bea
   & & {2 \over 3} \kappa_{,i}
       + {c \over {\cal N} ( 1 + 2 \varphi )}
       \left( {1 \over 2} \Delta \chi_i
       + {1 \over 6} \chi^k_{\;\;,ki} \right)
       + {8 \pi G \over c^4} \left( \mu + p \right)
       \gamma^2 v_i
   \nonumber \\
   & & \qquad
       =
       {c \over {\cal N} ( 1 + 2 \varphi)}
       \bigg\{
       \left( {{\cal N}_{,j} \over {\cal N}}
       - {\varphi_{,j} \over 1 + 2 \varphi} \right)
       \left[ {1 \over 2} \left( \chi^{j}_{\;\;,i} + \chi_i^{\;,j} \right)
       - {1 \over 3} \delta^j_i \chi^k_{\;\;,k} \right]
       - {\varphi^{,j} \over (1 + 2 \varphi)^2}
       \left( \chi_{i} \varphi_{,j}
       + {1 \over 3} \chi_{j} \varphi_{,i} \right)
   \nonumber \\
   & & \qquad
       + {{\cal N} \over 1 + 2 \varphi} \nabla_j
       \left[ {1 \over {\cal N}} \left(
       \chi^{j} \varphi_{,i}
       + \chi_{i} \varphi^{,j}
       - {2 \over 3} \delta^j_i \chi^{k} \varphi_{,k} \right) \right]
       \bigg\}
       - {1 \over 1 + 2 \varphi}
       {8 \pi G \over c^4} \Pi_{ij}
       v^j.
   \label{eq3}
\eea
Trace of ADM propagation:
\bea
   & & - {4 \pi G \over c^2} \left( \mu + 3 p \right)
       + {1 \over {\cal N}} \dot \kappa
       + {c^2 \Delta {\cal N} \over {\cal N} (1 + 2 \varphi)}
       = {1 \over 3} \kappa^2
       + {8 \pi G \over c^2} \left( \mu + p \right)
       \left( \gamma^2 - 1 \right)
   \nonumber \\
   & & \qquad
       - {c \over {\cal N} (1 + 2 \varphi)} \left(
       \chi^{i} \kappa_{,i}
       + c {\varphi^{,i} {\cal N}_{,i} \over 1 + 2 \varphi} \right)
       + c^2 \overline{K}^i_j \overline{K}^j_i
       + {1 \over 1 + 2 \varphi}
       {4 \pi G \over c^2} \left( \Pi^i_i
       + {1 \over 1 + 2 \varphi} \Pi_{ij}
       {v^i v^j \over c^2} \right).
   \label{eq4}
\eea
Tracefree ADM propagation:
\bea
   & & \left( {1 \over {\cal N}} {\partial \over \partial t}
       - \kappa
       + {c \chi^{k} \over {\cal N} (1 + 2 \varphi)} \nabla_k \right)
       \bigg\{ {c \over {\cal N} (1 + 2 \varphi)}
       \left[
       {1 \over 2} \left( \chi^i_{\;\;,j} + \chi_j^{\;,i} \right)
       - {1 \over 3} \delta^i_j \chi^k_{\;\;,k}
       - {1 \over 1 + 2 \varphi} \left( \chi^{i} \varphi_{,j}
       + \chi_{j} \varphi^{,i}
       - {2 \over 3} \delta^i_j \chi^{k} \varphi_{,k} \right)
       \right] \bigg\}
   \nonumber \\
   & & \qquad
       - {c^2 \over ( 1 + 2 \varphi)}
       \left[ {1 \over 1 + 2 \varphi}
       \left( \nabla^i \nabla_j - {1 \over 3} \delta^i_j \Delta \right) \varphi
       + {1 \over {\cal N}}
       \left( \nabla^i \nabla_j - {1 \over 3} \delta^i_j \Delta \right) {\cal N} \right]
   \nonumber \\
   & & \qquad
       =
       {8 \pi G \over c^2} \left( \mu +  p \right)
       \left[ {\gamma^2 v^i v_j \over c^2 (1 + 2 \varphi)}
       - {1 \over 3} \delta^i_j \left( \gamma^2 - 1 \right)
       \right]
       + {c^2 \over {\cal N}^2 (1 + 2 \varphi)^2}
       \bigg[
       {1 \over 2} \left( \chi^{i,k} \chi_{j,k}
       - \chi^{k,i} \chi_{k,j} \right)
   \nonumber \\
   & & \qquad
       + {1 \over 1 + 2 \varphi} \left(
       \chi^{k,i} \chi_k \varphi_{,j}
       - \chi^{i,k} \chi_j \varphi_{,k}
       + \chi_{k,j} \chi^k \varphi^{,i}
       - \chi_{j,k} \chi^i \varphi^{,k} \right)
       + {2 \over (1 + 2 \varphi)^2} \left(
       \chi^{i} \chi_{j} \varphi^{,k} \varphi_{,k}
       - \chi^{k} \chi_{k} \varphi^{,i} \varphi_{,j} \right) \bigg]
   \nonumber \\
   & & \qquad
       - {c^2 \over (1 + 2 \varphi)^2}
       \left[ {3 \over 1 + 2 \varphi}
       \left( \varphi^{,i} \varphi_{,j}
       - {1 \over 3} \delta^i_j \varphi^{,k} \varphi_{,k} \right)
       + {1 \over {\cal N}} \left(
       \varphi^{,i} {\cal N}_{,j}
       + \varphi_{,j} {\cal N}^{,i}
       - {2 \over 3} \delta^i_j \varphi^{,k} {\cal N}_{,k} \right) \right]
   \nonumber \\
   & & \qquad
       + {1 \over 1 + 2 \varphi}
       {8 \pi G \over c^2} \left( \Pi^i_j
       - {1 \over 3} \delta^i_j \Pi^k_k \right),
   \label{eq5}
\eea
ADM energy conservation:
\bea
   & & {1 \over {\cal N}} \left[ \mu
       + \left( \mu + p \right)
       \left( \gamma^2 - 1 \right) \right]^{\displaystyle\cdot}
       + {c \over {\cal N}} {\chi^i \over 1 + 2 \varphi}
       \left[ \mu
       + \left( \mu + p \right)
       \left( \gamma^2 - 1 \right) \right]_{,i}
   \nonumber \\
   & & \qquad
       - \left( \mu + p \right)
       {1 \over 3} \left( 4 \gamma^2 - 1 \right) \kappa
       + \left( { \mu + p \over 1 + 2 \varphi } \gamma^2 v^i \right)_{,i}
       +\left( {3 \varphi_{,i} \over 1 + 2 \varphi}
       + 2 {{\cal N}_{,i} \over {\cal N}} \right)
       {\mu + p \over 1 + 2 \varphi }
       \gamma^2 v^i
   \nonumber \\
   & & \qquad
       + {\gamma^2 (\mu + p)
       \over c {\cal N} ( 1 + 2 \varphi)^2}
       \left[ \chi^{i,j} v_{i} v_{j}
       - {1 \over 3} \chi^j_{\;\;,j} v^i v_{i}
       - {2 \over 1 + 2 \varphi}
       \left( v^i v^j \chi_i \varphi_{,j}
       - {1 \over 3} v^i v_{i} \chi^j \varphi_{,j} \right)
       \right]
       = - \Pi^{\rm (ADM)}.
   \label{eq6-ADM}
\eea
ADM momentum conservation:
\bea
   & & \left( {1 \over {\cal N}}
       {\partial \over \partial t}
       -\kappa \right)
       \left[ (\mu + p ) \gamma^2
       v_{i} \right]
       + {c \over {\cal N}} {\chi^j \over 1 + 2 \varphi}
       \left[ (\mu + p)
       \gamma^2 v_{i} \right]_{,j}
       + c^2 p_{,i}
       + c^2 \left( \mu + p \right)
       {{\cal N}_{,i} \over {\cal N}}
       + \left( {\mu + p \over 1 + 2 \varphi}
       \gamma^2 v^j v_{i} \right)_{,j}
   \nonumber \\
   & & \qquad
       + {c \over {\cal N}} \left( {\chi^j \over 1 + 2 \varphi} \right)_{,i}
       \left( \mu + p \right)
       \gamma^2 v_{j}
       + {\mu + p \over 1 + 2 \varphi}
       \gamma^2 v^j
       \left[ {1 \over 1 + 2 \varphi} (3 v_{i} \varphi_{,j}
       - v_{j} \varphi_{,i} )
       + {1 \over {\cal N}} ( v_{i} {\cal N}_{,j}
       + v_{j} {\cal N}_{,i} ) \right]
       = - c \Pi_i^{\rm (ADM)}.
   \label{eq7-ADM}
\eea
Covariant energy conservation:
\bea
   & & \left[ {\partial \over \partial t}
       + {1 \over 1 + 2 \varphi}
       \left( {\cal N} v^i + c \chi^i \right)
       \nabla_i \right] \mu
       + \left( \mu +  p \right)
       \bigg\{ - {\cal N} \kappa
       + { ( {\cal N} v^i )_{,i} \over 1 + 2 \varphi}
       +{ {\cal N} v^i \varphi_{,i} \over ( 1 + 2 \varphi )^2}
   \nonumber \\
   & & \qquad
       + {1 \over \gamma}
       \left[ {\partial \over \partial t}
       + {1 \over 1 + 2 \varphi }
       \left( {\cal N} v^i + c \chi^i \right)
       \nabla_i \right] \gamma
       \bigg\}
       = - \Pi^{\rm (ADM)}
       + {v^i \over c (1 + 2 \varphi)} \Pi_i^{\rm (ADM)}.
   \label{eq6-cov}
\eea
Covariant momentum conservation:
\bea
   & & {\partial \over \partial t}
       \left( \gamma v_{i} \right)
       + {1 \over 1 + 2 \varphi}
       \left( {\cal N} v^k + c \chi^k \right)
       \nabla_k \left( \gamma  v_{i} \right)
       + c^2 \gamma {\cal N}_{,i}
       + {1 - \gamma^2 \over \gamma}
       {c^2 {\cal N} \varphi_{,i} \over 1 + 2 \varphi}
       + c \gamma v^k
       \nabla_i \left( {\chi_k \over 1 + 2 \varphi} \right)
   \nonumber \\
   & & \qquad
       + {1 \over \mu + p}
       \left\{ c^2 {{\cal N} \over  \gamma} p_{,i}
       + \gamma v_{i}
       \left[ {\partial \over \partial t}
       + {1 \over 1 + 2 \varphi}
       \left( {\cal N} v^k
       + c \chi^k \right) \nabla_k \right]
       p \right\}
   \nonumber \\
   & & \qquad
       = {c {\cal N} \over (\mu + p) \gamma}
       \left[ {\gamma^2 v_i \over c}  \Pi^{\rm (ADM)}
       - \left( \delta^j_i
       + {\gamma^2 v_i v^j \over c^2 (1 + 2 \varphi)}
       \right) \Pi_j^{\rm (ADM)} \right].
   \label{eq7-cov}
\eea
We have
\bea
   & &
       \overline{K}^i_j \overline{K}^j_i
       = {1 \over {\cal N}^2 (1 + 2 \varphi)^2}
       \bigg\{
       {1 \over 2} \chi^{i,j} \left( \chi_{i,j} + \chi_{j,i} \right)
       - {1 \over 3} \chi^i_{\;\;,i} \chi^j_{\;\;,j}
       - {4 \over 1 + 2 \varphi} \left[
       {1 \over 2} \chi^i \varphi^{,j} \left(
       \chi_{i,j} + \chi_{j,i} \right)
       - {1 \over 3} \chi^i_{\;\;,i} \chi^j \varphi_{,j} \right]
   \nonumber \\
   & & \qquad
       + {2 \over (1 + 2 \varphi)^2} \left(
       \chi^{i} \chi_{i} \varphi^{,j} \varphi_{,j}
       + {1 \over 3} \chi^i \chi^j \varphi_{,i} \varphi_{,j} \right) \bigg\},
   \label{K-bar-eq}
\eea
for the tracefree part of extrinsic curvature $\overline{K}^i_j$,
and
\bea
   & &
       \Pi^{\rm (ADM)}
       \equiv \left( {1 \over {\cal N}} {\partial \over \partial t}
       + {c \chi^k \over {\cal N} (1 + 2 \varphi)} \nabla_k
       - {4 \over 3} \kappa \right)
       \left( { v^i v^j \over c^2 ( 1 + 2 \varphi )^2} \Pi_{ij} \right)
   \nonumber \\
   & & \qquad
       + {c \over {\cal N} (1 + 2 \varphi)^2}
       \left( \chi^i_{\;,j} - {2 \varphi^{,i} \chi_j \over 1 + 2 \varphi} \right)
       \left( \Pi^j_i - {1 \over 3} \delta^j_i \Pi^k_k \right)
       +\left( \nabla_i + 2 {{\cal N}_{,i} \over {\cal N}}
       + {3 \varphi_{,i} \over 1 + 2 \varphi} \right)
       \left( {v^j \over ( 1 + 2 \varphi )^2} \Pi^i_j \right),
   \label{Pi-ADM} \\
   & &
       \Pi_i^{\rm (ADM)}
       \equiv \left( {1 \over {\cal N}} {\partial \over \partial t}
       + {c \chi^k \over {\cal N} (1 + 2 \varphi)} \nabla_k
       - \kappa \right)
       \left( {v^j \over c ( 1 + 2 \varphi )} \Pi_{ij} \right)
       + {1 \over {\cal N} (1 + 2 \varphi)}
       \left( {\chi^j \over 1 + 2 \varphi} \right)_{,i}
       {v^k} \Pi_{jk}
   \nonumber \\
   & & \qquad
       + {c {\cal N}_{,i} \over {\cal N} (1 + 2 \varphi)^2}
       {v^j v^k \over c^2} \Pi_{jk}
       + \left( \nabla_j + 3 {\varphi_{,j} \over 1 + 2 \varphi}
       + {{\cal N}_{,j} \over {\cal N}} \right)
       \left( {c \over 1 + 2 \varphi} \Pi^j_i \right)
       - {c \varphi_{,i} \over (1 + 2 \varphi)^2} \Pi^j_j.
   \label{Pi-ADM-i}
\eea
Here we add the continuity equation, $(\overline \varrho u^c)_{;c} = 0$,
\bea
   & & \left[ {\partial \over \partial t}
       + {1 \over 1 + 2 \varphi}
       \left( {\cal N} v^i + c \chi^i \right)
       \nabla_i \right] \overline \varrho
       + \overline \varrho
       \left\{ - {\cal N} \kappa
       + { ( {\cal N} v^i )_{,i} \over 1 + 2 \varphi}
       +{ {\cal N} v^i \varphi_{,i} \over ( 1 + 2 \varphi )^2}
       + {1 \over \gamma}
       \left[ {\partial \over \partial t}
       + {1 \over 1 + 2 \varphi }
       \left( {\cal N} v^i + c \chi^i \right)
       \nabla_i \right] \gamma
       \right\}
       = 0.
   \nonumber \\
   \label{eq0}
\eea
This can be derived using Equations (C2), (C3), (D10), (D11) in HNa.

\end{widetext}

%
%


\end{document}